\documentclass[sigconf]{acmart}

\usepackage{hyperref} 
\usepackage{graphicx}
\usepackage{amsmath}
\usepackage{amsthm}
\usepackage{amsfonts}
\usepackage{subcaption}
\usepackage{balance}
\usepackage[ruled,vlined]{algorithm2e}
\usepackage{algpseudocode}
\usepackage{booktabs, multirow, tabularx}
\usepackage{xspace}
\usepackage{enumitem}
\usepackage[a-2b]{pdfx}

\newtheorem{defn}{Definition}

\newtheorem{problem}{Problem}

\newcommand{\squishlist}{
 \begin{list}{$\bullet$}
  {  \setlength{\itemsep}{0pt}
     \setlength{\parsep}{3pt}
     \setlength{\topsep}{3pt}
     \setlength{\partopsep}{0pt}
     \setlength{\leftmargin}{2em}
     \setlength{\labelwidth}{1.5em}
     \setlength{\labelsep}{0.5em}
} }
\newcommand{\squishlisttight}{
 \begin{list}{$\bullet$}
  { \setlength{\itemsep}{0pt}
    \setlength{\parsep}{0pt}
    \setlength{\topsep}{0pt}
    \setlength{\partopsep}{0pt}
    \setlength{\leftmargin}{2em}
    \setlength{\labelwidth}{1.5em}
    \setlength{\labelsep}{0.5em}
} }

\newcommand{\squishdesc}{
 \begin{list}{}
  {  \setlength{\itemsep}{0pt}
     \setlength{\parsep}{3pt}
     \setlength{\topsep}{3pt}
     \setlength{\partopsep}{0pt}
     \setlength{\leftmargin}{1em}
     \setlength{\labelwidth}{1.5em}
     \setlength{\labelsep}{0.5em}
} }

\newcommand{\squishend}{
  \end{list}
}

\newcommand{\eat}[1]{}

\newcommand{\kw}[1]{{\ensuremath {\mathsf{#1}}}\xspace}

\newcommand{\stitle}[1]{\vspace*{0.5em}\noindent{\bf #1}}

\newcommand{\eetitle}[1]{\vspace{0.8ex}\noindent{\em\underline{#1}}}

\newcounter{ccc}

\newcommand\redout{\bgroup\markoverwith
{\textcolor{red}{\rule[.5ex]{2pt}{2pt}}}\ULon}

\newcommand{\pg}{\kw{PG}}

\newcommand{\anns}{\kw{ANNS}}

\newcommand{\gate}{\kw{GATE}}

\AtBeginDocument{%
  }
    
\copyrightyear{2025}
\acmYear{2025}
\setcopyright{acmlicensed}\acmConference[KDD '25]{Proceedings of the 31st ACM SIGKDD Conference on Knowledge Discovery and Data Mining V.2}{August 3--7, 2025}{Toronto, ON, Canada}
\acmBooktitle{Proceedings of the 31st ACM SIGKDD Conference on Knowledge Discovery and Data Mining V.2 (KDD '25), August 3--7, 2025, Toronto, ON, Canada}
\acmDOI{10.1145/3711896.3736930}
\acmISBN{979-8-4007-1454-2/2025/08}

\begin{document}

\title{Empowering Graph-based Approximate Nearest Neighbor Search with Adaptive Awareness Capabilities}


\author{Jiancheng Ruan}
\authornote{Both authors contributed equally to this research.}
\orcid{0009-0008-1013-4690}
\affiliation{%
  \institution{Zhejiang University}
  \city{Hangzhou}
  \country{China}
}
\email{jiancheng.ruan@zju.edu.cn}

\author{Tingyang Chen}
\authornotemark[1]
\orcid{0009-0008-5635-9326}
\affiliation{%
  \institution{Zhejiang University}
  \city{Hangzhou}
  \country{China}
}
\email{chenty@zju.edu.cn}

\author{Renchi Yang}
\orcid{0000-0002-7284-3096}
\affiliation{%
  \institution{Hong Kong Baptist University
}
  \city{Hong Kong SAR}
  \country{China}
}
\email{renchi@hkbu.edu.hk}

\author{Xiangyu Ke}
\authornote{Corresponding author.}
\orcid{0000-0001-8082-7398}
\affiliation{%
    \institution{Zhejiang University, Zhejiang Key Laboratory of Big Data Intelligent Computing}
  \city{Hangzhou}
  \country{China}
}
\email{xiangyu.ke@zju.edu.cn}

\author{Yunjun Gao}
\orcid{0000-0003-3816-8450}
\affiliation{%
    \institution{Zhejiang University, Zhejiang Key Laboratory of Big Data Intelligent Computing}
  \city{Hangzhou}
  \country{China}
}
\email{gaoyj@zju.edu.cn}

\renewcommand{\shortauthors}{Jiancheng Ruan, Tingyang Chen, Renchi Yang, Xiangyu Ke, and Yunjun Gao}

\begin{abstract}
{\em Approximate Nearest Neighbor Search} (\anns) in high-dimensional spaces finds extensive applications in databases, information retrieval, recommender systems, etc.
While graph-based methods have emerged as the leading solution for \anns due to their superior query performance, they still face several challenges, such as struggling with local optima and redundant computations. 
These issues arise because existing methods (i) fail to fully exploit the topological information underlying the proximity graph $G$, and (ii) suffer from severe distribution mismatches between the base data and queries in practice.

To this end, this paper proposes \gate, high-tier proximity \underline{\textbf{G}}raph with \underline{\textbf{A}}daptive \underline{\textbf{T}}opology and Query Awar\underline{\textbf{E}}ness, as a lightweight and adaptive module atop the graph-based indexes to accelerate \anns. 
Specifically, \gate formulates the critical problem to identify an {\em optimal} entry point in the proximity graph for a given query, facilitating faster online search.
By leveraging the inherent clusterability of high-dimensional data, \gate first extracts a small set of {\em hub nodes} $\mathcal{V}$ as candidate entry points. 
Then, resorting to a contrastive learning-based two-tower model, \gate encodes both the structural semantics underlying $G$ and the query-relevant features into the latent representations of these hub nodes $\mathcal{V}$. 
A navigation graph index on $\mathcal{V}$ is further constructed to minimize the model inference overhead. 
Extensive experiments demonstrate that \gate achieves a 1.2-2.0$\times$ speed-up in query performance compared to state-of-the-art graph-based indexes.

\end{abstract}




\maketitle

\newcommand\kddavailabilityurl{https://doi.org/10.5281/zenodo.15523071}

\ifdefempty{\kddavailabilityurl}{}{
\begingroup\small\noindent\raggedright\textbf{KDD Availability Link:}\\
The source code of this paper has been made publicly available at \url{\kddavailabilityurl}.
\endgroup
}

\section{Introduction}
\label{sec:intro}

With the advent of deep learning, data of various modalities, such as text, audio, and images, can be transformed into high-dimensional embedding vectors with rich semantics, which has profoundly revolutionized how we represent and store data and led to the rise of vector databases~\cite{pan2023survey,guo2022manu}.
{\em Approximate Nearest Neighbor Search} (\anns)~\cite{li2019approximate}, which seeks to retrieve from a vector database $\mathcal{D} \subset \mathbb{R}^d$ the vectors that are approximately closest to the given query vector $q$ in polynomial time complexity, are acting as building blocks to underpin extensive practical applications and services, including information retrieval \cite{petitjean2014dynamic}, recommendation systems \cite{yi2019sampling, huang2019fibinet}, and Retrieval-Augmented Generation (\textsf{RAG}) \cite{lewis2020retrieval}.
Amid the plethora of acceleration technologies for \anns~\cite{fu2019fast, fu2016efanna, fu2021high, harwood2016fanng, li2019approximate}, graph-based techniques~\cite{wang2021comprehensive} have garnered significant attention from both academia and industry due to their superior query performance and high scalability, particularly for large-scale vector databases~\cite{pan2023survey}.

In the graph-based methodology, a vector database $\mathcal{D} \subset \mathbb{R}^d$ is often indexed as a {\em proximity graph} $G$, wherein each vector $p \in \mathcal{D}$ is regarded as a node, while edges connecting vectors are formed based on their proximity (e.g., Euclidean distance). 
In the query phase, a standard graph traversal will be conducted over $G$ to iteratively
search nodes with minimum distances to the query node~\cite{wang2021comprehensive} (detailed in Algorithm~\ref{alg:gnns}).
Over the past few years, considerable efforts have been devoted to optimizing the index structure for better graph navigability and richer semantic expressiveness~\cite{dong2011efficient, bratic2018nn, zhao2018k}. 
Despite the remarkable progress made in improving the query efficiency, the majority of previous works~\cite{ge2013optimized, jegou2010product, norouzi2013cartesian, zhang2019grip,abdi2010principal, munoz2019hierarchical, fu2019fast, fu2016efanna, fu2021high} focus on alleviating the impact of vector dimension $d$ and average degree $r$ in the query complexity dominated by $O(dr\ell)$, while the reduction of another complexity factor, i.e., the search path length $\ell$, is as of yet under-explored.
The reason is that 
$\ell$ is jointly affected by the index structure, search strategy, query distribution, and data characteristics, rendering it inherently challenging to optimize~\cite{wang2021comprehensive}. 
In particular, existing methods suffer from two major limitations pertaining to the search path length $\ell$. \

\stitle{{Limitation I: Graph Topology Unawareness.}} Several recent attempts~\cite{zhao2023towards,lu2021hvs, malkov2014approximate,malkov2018efficient,chen2023finger} for shortening search paths typically rely on establishing ``long-range'' connections within proximity graphs, which are often constructed in multi-layer structures. 
These approaches identify a set $\mathcal{S}$ of representative data points (e.g., via navigation small-world~\cite{malkov2014approximate} or clustering-based paradigms~\cite{lu2021hvs}) to guide the search, where the traversal direction is {\em solely determined by the distance} between the query vector $q$ and data points in $\mathcal{S}$. 
However, such a search scheme is inefficient as it fails to harness the topological connectivity of $G$ and tends to get stuck in a local region of $G$ whose navigability is not optimized for the target results $\mathcal{T}$ of $q$, particularly in the frequently observed case where the shortest paths from $\mathcal{S}$ to $\mathcal{T}$ are truncated~\cite{fu2019fast}. 

Despite the great potential of graph topology in reducing search path length $\ell$, the effective exploitation of it is non-trivial and still remains challenging.
In particular, graph indexes are often based on well-established theoretical graphs, e.g., \textsf{Delaunay graph}\cite{fortune1992voronoi} and \textsf{RNG}\cite{toussaint1980relative}, each of which presents distinct topological structures specially catered for various retrieval scenarios. On the 
 other hand, optimization techniques that rely on specific graph structures often lack generalization capabilities~\cite{zhang2022grasp,baranchuk2019learning}. 

\stitle{{Limitation II: Complex Query Distributions.}}
In real-world applications, query vector $q$ and target $\mathcal{T}$ are usually of different modalities, e.g., textual queries and images for retrieval, which engenders the issue of {\em distributional mismatches}.
On top of that, real query vectors are usually from diverse sources and exhibit significant heterogeneity and variability. The search path lengths thus can be highly variable, even their search source points on $G$ are the same.
As an aftermath, the distribution mismatch between the graph index and queries is exacerbated, which induces considerable computational redundancy and increases the risk of returning suboptimal results \cite{wang2020note}.
These problems necessitate an \anns framework that should be not only aware of the structure of $G$ but also adaptive to {\em any} possible queries. Unfortunately, in real-world scenarios such as e-commerce search, graph-based \anns typically follows a ``build-then-search'' paradigm, in which the queries are unknown in the build phase~\cite{chen2024roargraph}.
Notice that it is also infeasible to specialize optimizations for a query distribution since queries can come from multiple modalities in practice, making the design of an adaptive \anns framework a formidable challenge.

In response, this paper presents \gate, a novel high-tier proximity \underline{\textbf{G}}raph with \underline{\textbf{A}}daptive \underline{\textbf{T}}opology and Query Awar\underline{\textbf{E}}ness that enables adaptive navigation on graph indexes for \anns.
The core innovation of \gate lies in the introduction of {critical navigation points} $\mathcal{V}$, henceforth referred to as {\em hub nodes}, which are identified based on the clusterability of high-dimensional data and can be served as {\em hubs} for traversal over $G$ (\S\ref{sec:view-nodes}).

At a high level, we draw inspiration from the popular {\em two-tower model}~\cite{yi2019sampling,huang2019fibinet} in recommendation systems, wherein users and items are first jointly encoded before pinpointing matched items for each user.
In the same vein, we innovatively frame our task as ``recommending'' the optimal {\em entry point} from $\mathcal{V}$ that leads to shortest search paths on $G$ to the query $q$ through the two-tower model.
The linchpin, therefore, turns to refine the latent representations of the hub nodes in $\mathcal{V}$ by incorporating topological semantics in $G$ and query-specific features that are conducive to fast search.

Specifically, we first develop a subgraph sampling technique to extract from $G$ the key structural patterns pertinent to nodes in $\mathcal{V}$ (\S\ref{sec:two-pro}). 
The sampled subgraphs are later encoded into structure embeddings and further injected into the data vectors of the hub nodes to generate their augmented latent representations. 
To achieve query awareness, we resort to contrastive learning along with the two-tower model to further refine the latent representations of the hub nodes, in which positive and negative samples for each hub node are taken from historical queries to capture hidden connections between the base data and queries.
Lastly, an additional navigation graph \gate is then constructed by establishing connections between close hub nodes, so as to prevent excessive model inferences during the search and facilitate the efficient graph traversal (\S\ref{sec:high-tier}).
Building on the foregoing design and optimizations, \gate is able to enhance \anns efficiency under any existing graph indexes without requiring any alteration to their structures, and hence, can be employed as a lightweight plug-and-play module to bolster graph-based \anns.
Our extensive experiments comparing \gate against five recent graph-based \anns methods on five benchmark datasets demonstrate that \gate can achieve a notable speedup of 1.2-2$\times$ in query performance, at a small cost of the additional index and model training (\S\ref{sec:Exp}).

\vspace{-3mm}
\section{Preliminaries}
\label{sec:prelim}
This section provides an overview of {\em Approximate Nearest Neighbor Search} (\anns), focusing on the use of proximity graphs for indexing and efficient search. 
We then discuss the {\em two-tower model} and its alignment with \anns for optimizing search performance. 
Relevant notation is summarized in Table~\ref{tab:notation}.

\begin{table}[tb!]
\caption{Table of notations}
\label{tab:notation}
\vspace{-3mm}
\begin{tabular}{c|c}
\textbf{Symbol} & \textbf{Meaning} \\
\hline \hline
$\mathbb{R}^d$ & The euclidean space in $d$-dimension \\
\hline 
$G$ = $(V, E)$ & A proximity graph with nodes $V$ and edges $E$ \\
\hline 
$\delta(\cdot,\cdot)$ & $l_2$ distance between two vectors \\
\hline 
$S(\cdot,\cdot)$ & Cosine similarity between two vectors \\
\hline 
$\lVert \cdot \rVert$ & The $l_2$ norm of a vector \\
\hline 
$\mathcal{V}$ & A set of hub nodes \\
\hline 
$\mathcal{Q}$ & A set of queries \\
\hline 
\end{tabular}
\vspace{-1ex}
\end{table}

\subsection{Background and Problem Definition}

\stitle{ANNS.} Numerous real-world tasks, such as information retrieval and question answering, can be conceptualized as exact nearest neighbor search problems. However, locating the exact nearest neighbor in large, high-dimensional datasets is widely acknowledged to be a time-consuming endeavor. 
Thus, this paper focuses on a more efficient and potentially more practical alternative: approximate nearest neighbor search (\anns).
\begin{problem}[$\epsilon$-Approximate Nearest Neighbor Search~\cite{indyk1998approximate}]
Given a set of vectors $\mathcal{D} \subset \mathbb{R}^d$ , a query vector $q \subset \mathbb{R}^d$, and an approximation ratio $\epsilon$, the $\epsilon$-ANNS aims to efficiently find a vector $p$ in $\mathcal{D}$ that satisfies $\delta(p, q) \le (1 + \epsilon)\delta(p^*, q)$, where $\delta(\cdot,\cdot)$ represents a distance metric in the Euclidean space and $p^*$ is the optimal result.
\end{problem}

The concept above can be easily extended to the \textbf{Approximate K Nearest Neighbor Search ({\sf AKNNS})} problem. 
For modeling and evaluation purposes, in practice, we use recall@$k$ instead of the predefined $\epsilon$ to assess the precision of {\sf AKNNS}. 
Specifically, suppose that $R'$ represents the $k$ vectors returned by a {\sf AKNNS} method, and $R$ denotes the ground truth set of $k$ nearest neighbors. 
The recall@$k$ metric is then computed as: 
\begin{equation}
\label{eq:recall}
    Recall@k = \frac{|R \cap R'|}{k}
\end{equation}

\eetitle{Remark.} For simplicity, we use \anns as the unified term in the following discussion unless there is ambiguity.

\stitle{Proximity Graph.} Proximity graphs (\pg) have been extensively studied as a foundation for graph clustering and query processing, owing to their ability to identify and utilize proximity relationships within datasets. 
Over time, various variants of \pg have been developed, each optimized for specific proximity criteria~\cite{fortune1992voronoi,toussaint1980relative,preparata2012computational}. 
More recently, \pg-based methods for nearest neighbor search have gained attention due to their impressive query performance and robust theoretical guarantees, formally defined as below:

\begin{defn}[Proximity Graph]
    Given a vector database $\mathcal{D}$, a proximity graph $G = (V, E)$ is constructed as:
    (1) Each node $v_i$ of $V$ represents a vector $p_i\in\mathcal{D}$, 
    and (2) Each edge $(u,v)\in E$ indicates that nodes $u$ and $v$ satisfy a predefined proximity criterion under a specific distance metric. 
\end{defn}

\stitle{Navigating Spreading-out Graph (\textsf{NSG})~\cite{fu2019fast}.} \textsf{NSG} is a popular proximity graph that approximates the Monotonic Relative Neighbor Graph (\textsf{MRNG})~\cite{fu2019fast}. It sorts the neighbors of nodes by Euclidean distance and prunes edges using the triangle inequality. Recognized as a state-of-the-art method for \anns, we select \textsf{NSG} as the representative proximity graph in this paper due to its efficiency and solid theoretical foundations. 
Note that our proposed method (\S\ref{sec:methods}) is a lightweight plugin that can seamlessly integrate with {\em any} underlying proximity graph.

\stitle{On-graph Searching.} Existing methods differ in index structures, but they generally follow the same greedy search strategy as shown in Algorithm~\ref{alg:gnns}~\cite{wang2021comprehensive}:  
Select an entry point (i.e., to start the retrieval), then iteratively navigate to the neighboring node with the shortest distance to the query node until the termination criteria are met. 
Hence, the search complexity is primarily determined by the number of steps (\(\ell\)) and the average out-degree (\(r\)) of the graph. 
However, current graph-based algorithms approximate the theoretical graph, potentially leading to local optima and redundant computations during search. To tackle these, \S\ref{sec:methods} explores how \gate avoids these local optima and reduces search length (\(\ell\)) to decrease time complexity for existing graph indexes.

\begin{algorithm}[t]

  \caption{\textsc{Greedy Search For Graphs}}
  \label{alg:gnns}
  \small
  \KwIn{Dataset $\mathcal{D}$, Graph $G$, query $q$, candidate set size $l_s$, result set size $k$. Euclidean distance function $\delta(,)$}
  \KwOut{Top $k$ result set $R$.}
  $R \gets \emptyset$; $Q \gets \emptyset$;\\
  $P \gets$ random sample $l_s$ nodes from $G$; \\
  \For{each node $p$ in $P$}{
    $Q$.add($p$,$\delta(p, q)$);
  }
  $Q$.make\_min\_heap(); $R$.init\_max\_heap() \\ 
  \While{$Q$.size()}{
    $p\gets Q.pop()[0]$; $R.insert((p, \delta(p, q)))$\\
    \If{visited($p$)}{continue;}
    $N_p \gets$ neighbors of $p$ in $G$ \\
    \For{each node $n$ in $N_p$}{
        $Q.insert((n,\delta(n,q)))$;
    }
    $Q.resize(l_s)$;$R.resize(k)$; \\
  }

  \textbf{return} $R$
\end{algorithm}

\subsection{Two-Tower Models}
Two-tower structured neural networks have recently become popular in recommendation systems~\cite{yi2019sampling, huang2019fibinet} since it can effectively model user-item feature connections.
This section presents the background and discusses their connection to \anns. 

\stitle{Feature Embeddings.} We consider a standard input for a two-tower model in which we have a set of queries (user needs) and items (products in the database). Queries and items are projected into a shared high-dimensional space by an advanced embedding model~\cite{yi2019sampling} to capture their respective semantic information. 
Thus, queries and items can be represented by feature embeddings $\{q_i\}_i^N \subset \mathbb{R}^d$ and $\{p_i\}_i^M \subset \mathbb{R}^d$ respectively~\cite{yi2019sampling}. 

\stitle{Training Objective.} The two-tower model calculates the similarity between a query and an item using cosine similarity, allowing interaction between two distinct embedding sets and capturing their underlying relationships. 
This enables the recommendation system to retrieve the most relevant subset of items for a given query. 
The similarity can be represented as two parameterized embedding functions: $\text{sim}(q, i) = \frac{f_q(q)^\top f_i(i)}{\|f_q(q)\| \|f_i(i)\|}$. Let $i$ represent the item, $q$ the query, and $f_q(\cdot)$ and $f_i(\cdot)$ denote the query and item embedding functions, respectively.

The training objective is to maximize the similarity between relevant query-item pairs while minimizing the similarity between irrelevant pairs. This can be formalized as:
\begin{equation}
    \label{eq:loss}
    \mathcal{L} = -\frac{1}{|\mathcal{P}(q)|} \sum_{(q, i^+) \in \mathcal{P}(q)} \log \frac{\exp(\text{sim}(q, i^+) / \tau)}{\sum_{i^- \in \mathcal{N}(q)} \exp(\text{sim}(q, i^-) / \tau)}
\end{equation}

Where $\mathcal{P}(q)$ is the training dataset containing query-positive item pairs $(q, i^+)$, $\mathcal{N}(q)$ is a set of negative items sampled for query $q$, $\tau$ is a temperature parameter~\cite{yi2019sampling}.

\stitle{Connection to \anns.} \anns aims to identify a set of neighbors in a database that are close to a given query in terms of Euclidean distance. 
While it differs from the two-tower model in the specific metric used, both share a similar motivation: to retrieve items that are similar to the query. 
However, the two-tower model employs learning techniques to capture the intrinsic relationships between queries and items, allowing for adaptive adjustments to their embeddings. 
This insight opens up opportunities to enhance \anns, which typically relies on fixed mathematical distances and overlooks the interaction between queries and items.

\section{Analyzing the Performance Bottlenecks in Graph-based Indexes}
\label{sec:analysis}
This section analyzes the limitations of current graph-based methods, focusing on two critical aspects that influence query performance. 
We begin with analyzing graph indexes and their key properties, followed by a summary that motivates the design of our approach presented in \S\ref{sec:methods}.

\stitle{Clusters in High-dimensional Space.} 
Our first analysis examines the clustering behavior of high-dimensional spaces, especially in modern embedding models, and its impact on \anns.  
In tasks like image retrieval~\cite{gillick2018end}, the goal is to retrieve the most similar items. 
To achieve this, training methods encourage similar items to cluster together in embedding space~\cite{radford2021learning}, which can enhance \anns performance for retrieval but leads to clustering effects~\cite{parsons2004subspace}.
Existing proximity graph algorithms sample only a portion of the points in $\mathcal{D}$ and connect similar points and apply pruning techniques on these points to create sparser graphs for better time-and-storage trade-off. These graphs are then used in a unified search algorithm (Algorithm~\ref{alg:gnns}) to process online queries.

\begin{figure}[tb!]
\vspace{1ex}
\centering
\centerline{\includegraphics[width=0.9\linewidth]{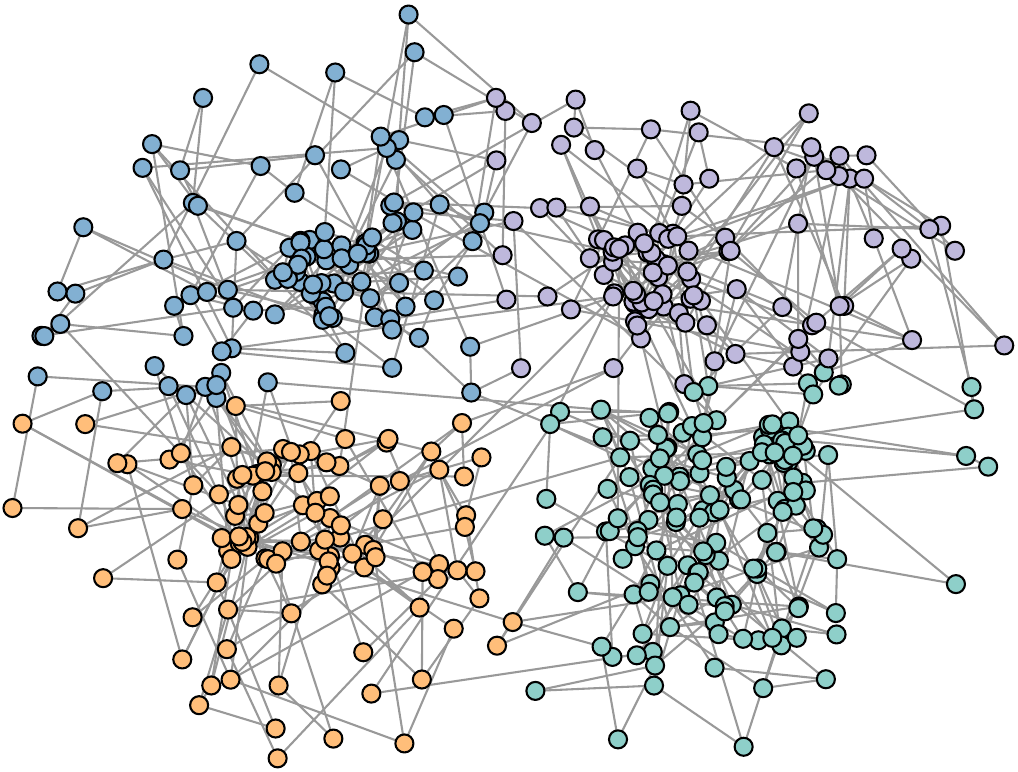}}
\vspace{-2ex}
\caption{Visualizing \textsf{NSG} on a sample of \textsf{Sift10M} points with t-SNE and K-means into 4 clusters~\cite{tang2016visualizing}.}
\vspace{-4ex}
\label{fig:dataset_view}
\end{figure}

We hypothesize that current proximity graph indexes, by relying on approximation, sacrifice structural information, leading to local optima and increased computational costs~\cite{yue2023routing}. Furthermore, clustering exacerbates these issues by: (1) reducing inter-cluster edge density, impairing connectivity (e.g., \textsf{ImageNet}~\cite{russakovsky2015imagenet} have more than 10,000 categories which naturally form a lot of clusters, where sparse inter-cluster edges hinder searching performance), and (2)  resulting in variable data densities and distinct similarity relationships across clusters. This leads to highly variable intra-cluster edge densities, misleading searches into suboptimal regions and increasing search path length (\(\ell\)) (see Figure~\ref{fig:dataset_view}).
This motivates the need to optimize query performance by considering the clustering properties of data, which we will address in \S\ref{sec:methods}.

\begin{figure}[tb!]
\vspace{1ex}
\centering
\centerline{\includegraphics[width=0.94\linewidth]{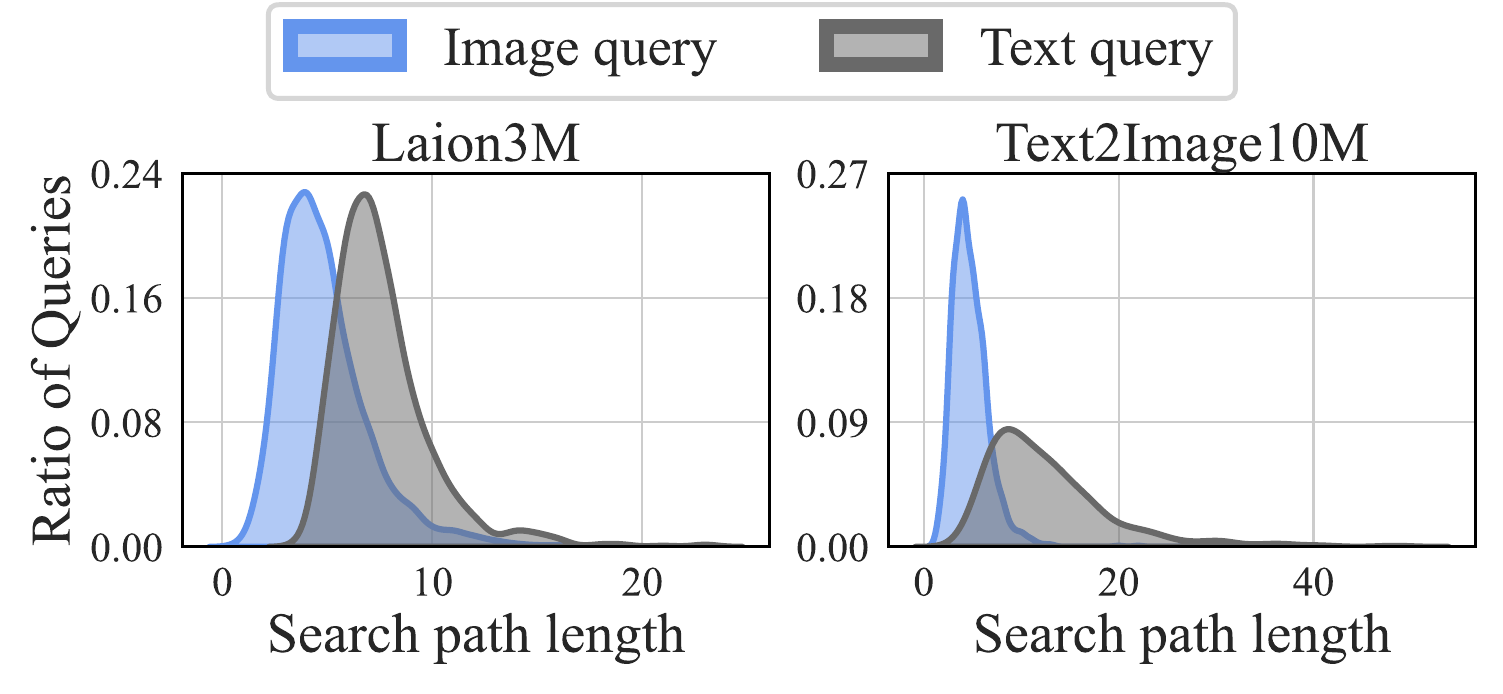}}
\vspace{-2ex}
\caption{Search path length (\(\ell\)) across different query types on \textsf{Laion3M} and \textsf{Text2image10M}.}
\vspace{-2ex}
\label{fig:variance_queries}
\end{figure}

\stitle{Variance in Queries.} 
Our second analysis highlights the importance of aligning query and base vector distributions. Existing proximity graph indexes primarily rely on distance metrics between database points. 
However, these methods often overlook the query distribution, which can differ from the base data in real-world scenarios. This mismatch can significantly affect search efficiency and lead to suboptimal performance. For example, supporting text queries against image databases is a common requirement. However, even within a shared latent space, indexes optimized for different modalities exhibit significant performance disparities~\cite{chen2024roargraph}.  While effectively supporting image-to-image retrieval, performance can degrade substantially for text-to-image  retrieval.  To quantify this modality gap, we conduct experiments on two benchmark datasets: \textsf{Laion3M} and \textsf{Text2image10M}, using 1,000 distinct queries per modality (text and image). Figure~\ref{fig:variance_queries} shows that achieving 95\% recall for top-1 nearest neighbor for text queries requires $2-5\times$ more search path length (\(\ell\)) for image queries. Since practical applications demand support for both query types but the image data itself may lack sufficient textual information. Therefore, there is room to improve retrieval efficiency by better-utilizing query distribution.

\stitle{Summary.} In conclusion, despite its advancements, graph-based methods still have inherent limitations that require further optimization. To further enhance their performance, this paper primarily focuses on addressing two key questions: (1) {\em How can the topological structure of graphs be leveraged to improve search performance?} (2) {\em How can query distributions be efficiently utilized to enhance the capability of graph-based methods in handling "unseen" queries?}

\section{Methodology}

\begin{figure*}[tb!]
\vspace{1ex}
\centering
\centerline{\includegraphics[width=\linewidth]{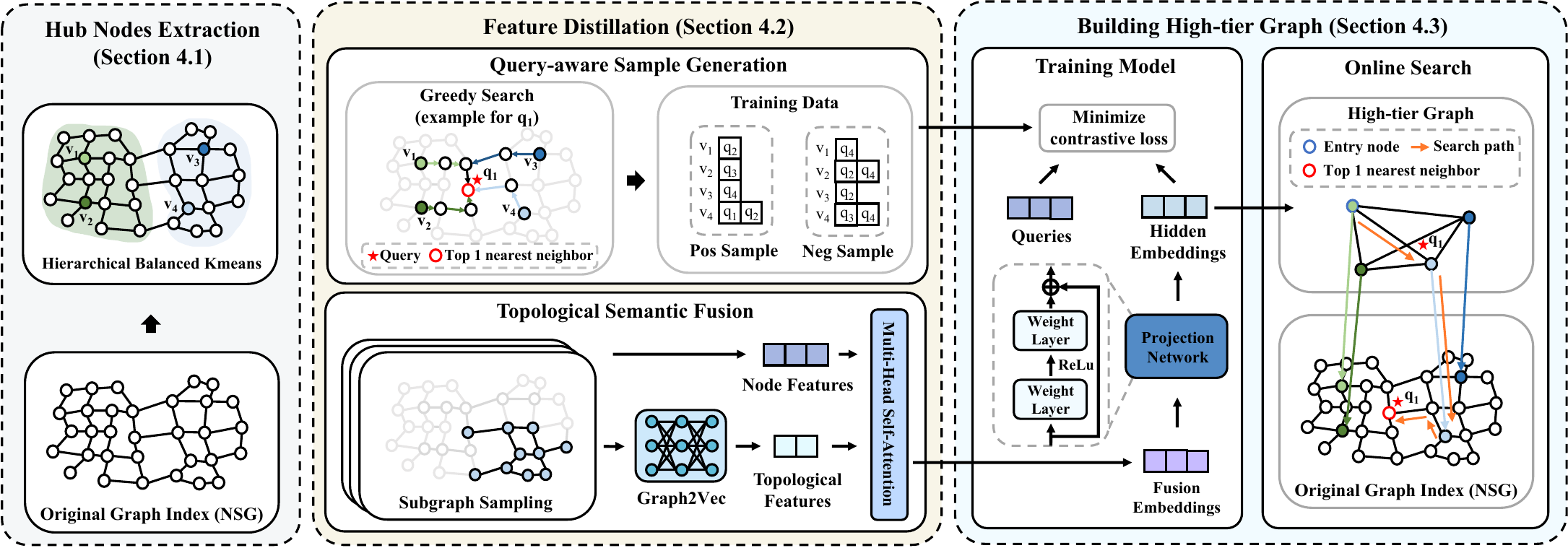}}
\vspace{-2ex}
\caption{A pipeline of \gate}
\vspace{-1ex}
\label{fig:pipline}
\end{figure*}

\label{sec:methods}

In this section, we present \gate, a novel method for building an adaptive high-tier navigation graph for arbitrary graph indexes. Our approach consists of three key steps: (1) identifying key navigation hubs (hub nodes) based on clusterability of high-dimensional data  (\S\ref{sec:view-nodes}); (2) extracting topological features and query distribution information based on these hub nodes (\S\ref{sec:two-pro}); and (3) constructing the adaptive high-tier navigation graph using a two-tower model (\S\ref{sec:high-tier}). While we use \textsf{NSG} as an example underlying graph in this paper, \gate can be applied to other graph indexes as well. The overall architecture of \gate is illustrated in Figure \ref{fig:pipline}.

\subsection{Hub Nodes Extraction}
\label{sec:view-nodes}
High-dimensional vectors often exhibit clustering properties~\cite{parsons2004subspace}, where vectors within the same cluster have high similarity, while connections between different clusters are relatively sparse.  This phenomenon leads to the following characteristics in existing proximity graphs: (1) dense intra-cluster connections and sparse inter-cluster links; (2) variations in topological structures among different clusters. Inspired by this, to better capture the topology of different clusters and enhance inter-cluster connectivity, we introduce hub nodes, which can be extracted as follows (Algorithm~\ref{alg:hbkm}):

\begin{defn}[Balanced Clusters]
    For a vector dataset $\mathcal{D} \subset \mathbb{R}^d$, a balanced $k$-clustering partitions $\mathcal{D}$ into $k$ disjoint clusters $\{C_{1}, C_{2}, \cdots, C_{k}\}$ such that the variance of cluster sizes is minimized: $\sum_{i=1}^{k} (|C_{i}| - \frac{|D|}{k})^2$.
\end{defn}

\begin{defn}[Hub Nodes]
    Given a vector dataset $\mathcal{D} \subset \mathbb{R}^d$, a node set size $n_c$, the hub nodes set $\mathcal{V}$ can be constructed as follows: (1) Partition $\mathcal{D}$ into $k$ balanced clusters: $\{C_{1}, \cdots, C_{k}\ = \text{HBKM}(\mathcal{D}, k, n_c)\}$ (Algorithm~\ref{alg:hbkm}). (2) For each cluster $C_i \in \{C_i, \dots,C_k \}$, compute its centroid $\mu_i = \frac{1}{|C_i|} \sum_{x \in C_i} x$ and select the nearest neighbor of $\mu_i$ as the hub node: $\mathcal{V}_i = \arg\min_{x \in C_i} \| x - \mu_i \|^2$. $\mathcal{V}$ is the union of hub nodes from each cluster: $\mathcal{V} = \bigcup_{i=1}^{k} \mathcal{V}_i$.
\end{defn}

The aforementioned hub node selection method offers two primary advantages: \textbf{(1) Increased Granularity:} Hierarchical clustering effectively captures the local structure of high-dimensional spaces from coarse to fine granularity, enabling the identification of significant hub points as cluster centers. \textbf{(2) Low Computational Cost:} The layered data partitioning reduces the computational cost per iteration, and the ability to process different clusters in parallel makes this method highly suitable for large-scale vector data. In summary, this approach achieves a balance between granularity and computational efficiency, making it well-suited for large-scale high-dimensional vector data.

Having identified suitable hub nodes, we now turn our attention to enriching them with topological and query awareness.
\setlength{\textfloatsep}{0pt} %
\begin{algorithm}[t]
\caption{Hierarchical Balanced K-Means (HBKM)}
\label{alg:hbkm}
\small
\KwIn{Dataset $\mathcal{D}$, branching factor $k$, target clusters $n_c$, penalty $\lambda$, max iterations $T$}
\KwOut{Balanced clusters $\{C_1, \dots, C_{n_c}\}$}
$l, l_{\text{max}}, C^{(0)}_1 \gets 0, \lceil \log_k{n_c} \rceil, \mathcal{D}$ \\
\While{$l < l_{\text{max}}$}{
    $l \gets l + 1$ \\
    \For{each cluster $C^{(l-1)}_i$}{
        $C^{(l)}_{ij} \gets \emptyset$ for $j = 1, \dots, k$ \\
        \For{$t = 1$ to $T$}{
            \For{each point $x \in C^{(l-1)}_i$}{
                $j = \arg\min_{j} \|x - \mu_{ij}\|^2 + \lambda \left( |C^{(l)}_{ij}| - \frac{|C^{(l-1)}_i|}{k} \right)^2$ \\
                $C^{(l)}_{ij} \gets C^{(l)}_{ij} \cup \{x\}$
            }
            \For{each $C^{(l)}_{ij}$}{
                $\mu^{(l)}_{ij} \gets \frac{1}{|C^{(l)}_{ij}|} \sum_{x \in C^{(l)}_{ij}} x$
            }
        }
    }
}
\textbf{return} clusters $\{C^{(l)}_1, \dots, C^{(l)}_{n_c}\}$.
\end{algorithm}
\setlength{\textfloatsep}{12pt plus 2pt minus 2pt}

We apply HKBM to partition the database $\mathcal{D}$ into a specified number of balanced clusters $C_1, C_2, \dots, C_{n_c}$, where $n_c$ denotes the total number of clusters. The notion of cluster balance implies minimizing the variance of cluster sizes, formalized by the objective function $\sum_{i=1}^{n_c}{\left(|C_{i}| - \frac{|\mathcal{D}|}{n_c}\right)^2}$. To address the cluster balance issue, we introduce a cluster size penalty term, $\lambda \left( |C^{(l)}_{i}| - \frac{|C^{(l-1)}|}{k} \right)^2$, into the cluster selection criterion of the k-means algorithm. Here, $k$ represents the number of partitions within a single execution of k-means, and $\lambda$ is a scaling factor controlling the penalty strength.

\subsection{Feature Distillation}
\label{sec:two-pro}
\stitle{Topology Features Extraction.}
The irregular structures of graphs pose a challenge to accurately capturing their topological features.  Recent research has shown that subgraph-based representation methods are effective at capturing graph topology~\cite{narayanan2017graph2vec}. However, the exponential number of possible subgraphs makes finding the optimal set of representative subgraphs computationally expensive. Here, we introduce an effective subgraph sampling technique for proximity graphs in \anns perspective based on hub nodes.

\begin{figure}[tb!]
\vspace{1ex}
\centering
\centerline{\includegraphics[width=\linewidth]{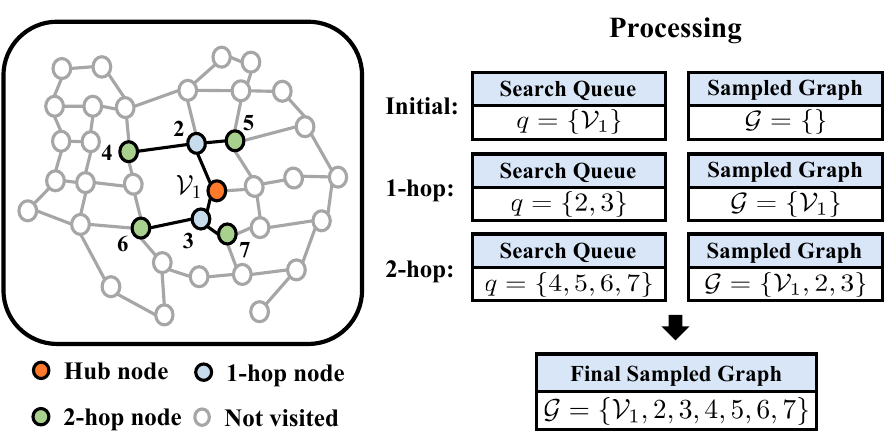}}
\vspace{-2ex}
\caption{An example of subgraph sampling for $\mathcal{V}_{1}$ ($h=2$).}
\vspace{-3ex}
\label{fig:subgraph_sampling}
\end{figure}

\eetitle{Subgraph Sampling.} Figure~\ref{fig:subgraph_sampling} shows an example of the subgraph sampling procedure of one hub node. Given a proximity graph $G$, a max hop $h$ and an initial visit queue $Q_v$ (initialized with $\mathcal{V}_i$), for each hub node $\mathcal{V}_i$, we construct a subgraph $G_{\mathcal{V}_i}$ by exploring the $h$-hop graph of $\mathcal{V}_i$ through a guided walk. The algorithm iteratively dequeues a node $v$ from $Q_v$. For the current node $v$, we sample $\lceil x/2 \rceil$ nearest neighbors and $\lfloor x/2 \rfloor$ farthest neighbors of $v$ on $G$ based on Euclidean distance. The number of neighbors to sample, $x$, is dynamically adjusted based on the node's degree and the overall graph's degree distribution: $x = \left\lceil \frac{\text{MinDegree}(G)}{\text{MaxDegree}(G)} \times \text{degree}(v) \right\rceil$.  These sampled neighbors and the corresponding edges are added to the subgraph $G_{\mathcal{V}_i}$. If any of the sampled neighbors are within $h$ hops of the originating hub node $\mathcal{V}_i$, they are enqueued for further exploration. This process is recursively applied to each newly visited node until the queue is empty. The motivations behind it are twofold: (1) This mixed sampling strategy, \textbf{considering short- and long-range neighbors, diversifies perceived local structures within $G$ and promotes inter-cluster exploration while preserving intra-cluster connections}. (2) By dynamically adjusting $x$, \textbf{it prevents over-sampling from dense nodes (high out-degree) and under-sampling from sparse nodes (low out-degree), ensuring balanced neighborhood expansion across various graph regions}.

Given the sampled subgraphs, we aim to create a unified representation of these irregular structures and effectively utilize the information they contain. To achieve this, we focus on learning the implicit semantic information within these irregular subgraphs and project them into fixed-dimensional topological features. This is accomplished by applying advanced graph embedding models, such as Graph2Vec~\cite{narayanan2017graph2vec}. Future advances in embedding methods can further refine these embeddings. Through this process, we have successfully captured topological information for each hub node. Each hub node $\mathcal{V}_{i}$ is now associated with a corresponding  topological feature $\mathcal{U}_{\mathcal{V}_{i}}$. In the following section, we will explore how to integrate query information.

\stitle{Query Information Enhancement.} To incorporate query information, we propose a sampling method designed to capture the interaction patterns between queries and hub nodes. Because the hub nodes are derived from different clusters, our goal is to enable each hub node to recognize the query types it can best serve.

\eetitle{Query-aware Sample Generation.} To achieve our goal of making hub nodes query-aware, we define positive and negative query samples for each hub node.

\begin{defn}[Positive Sample]
Given a proximity graph $G = (V, E)$, a set of queries $\mathcal{Q}$, and a hub node $\mathcal{V}_i$, let $H(q, \mathcal{V}_i)$ denote the hop count of the shortest path from $\mathcal{V}_i$ to the nearest neighbor of query $q \in \mathcal{Q}$ in $G$. A positive sample $q_{pos}$ for $\mathcal{V}_i$ is a query $q_{pos} \in \mathcal{Q}$ such that $H(q_{pos}, \mathcal{V}_i) \leq \min_{q \in \mathcal{Q}} H(q, \mathcal{V}_i) + t_{pos}$, where $t_{pos}$ is a pre-defined tolerance parameter.
\end{defn}

Negative samples are defined analogously. Query distributions typically exhibit relative consistency, allowing us to leverage multiple historical queries as $\mathcal{Q}$ to enhance sample generation. We maintain two sample queues for $\mathcal{V}_i$: a positive sample queue $\mathcal{Q}_i^+ = \{\}$ and a negative sample queue $\mathcal{Q}_i^- = \{\}$. In practice, we generate these samples by applying Algorithm~\ref{alg:gnns} on \textsf{NSG} for each query $q \in \mathcal{Q}$, and then assign them to the positive and negative sample queues of the corresponding hub node as follows:

\begin{itemize}[left=0pt, nosep]
    \item  For each query $q \in \mathcal{Q}$, we use each hub node $\mathcal{V}_i \in \mathcal{V}$ as a starting point and count the hop count from $\mathcal{V}_i$ to the top-1 result of $q$ on the \textsf{NSG} by employing Algorithm~\ref{alg:gnns}. If $q$ meets the criteria for a positive (or negative) sample, it is added to $\mathcal{V}_i$'s corresponding queue, indicating $\mathcal{V}_i$'s suitability (or unsuitability) as an entry point for $q$ in $G$. A threshold parameter $t_{pos}$ controls the balance between hard and easy positive samples.

\end{itemize}

Based on the that, Each hub node $\mathcal{V}_i$ gets two sample queues: a positive sample queue $\mathcal{Q}_i^+ = \{q_{pos_1}, q_{pos_2}, \dots, q_{pos_n}\}$ and a negative sample queue $\mathcal{Q}_i^- = \{q_{neg_1}, q_{neg_2}, \dots, q_{neg_n}\}$. The positive and negative samples for each hub node provide strong contrast, while samples from different hub nodes are distinctly dissimilar. This ensures that each hub node learns from both relevant and irrelevant queries, enhancing the hub node's discriminative power.  Having identified hub nodes and associated them with topological and query information, we next discuss how to integrate these elements to construct \gate.

\subsection{Building High-tier Graph}
\label{sec:high-tier}
This section introduces a contrastive learning-based two-tower model that integrates topological feature and query information to generate enhanced latent representations for the hub nodes.  By connecting these enhanced hub nodes, we construct \gate, an adaptive, high-tier navigation graph that exhibits both topology and query awareness.

\stitle{Model Architecture.} Our model comprises two primary modules: a Fusion Embedding Augmentation and a Projection Network.

\eetitle{Module I: Fusion Embedding Augmentation.}  Each hub node $\mathcal{V}_i$ is associated with a topological feature $\mathcal{U}_{\mathcal{V}_i}$ and a node feature $p_{\mathcal{V}_i}$. While $p_{\mathcal{V}_i}$ captures absolute spatial information, $\mathcal{U}_{\mathcal{V}_i}$ encodes relative positional relationships within the local graph structure~\cite{yin2022algorithm}. To effectively integrate these complementary aspects, we introduce a fusion module that augments the hub node feature $p_{\mathcal{V}_i}$ with topological feature from $\mathcal{U}_{\mathcal{V}_i}$. Specifically, we employ a multi-head attention mechanism to generate a fusion embedding $\mathcal{F}_{\mathcal{V}_i}$ for each hub node $\mathcal{V}_i$:
\begin{equation}
    \label{eq:fusion}
    \resizebox{0.92\linewidth}{!}{$
    \mathcal{F}_{\mathcal{V}_i} = \text{Concat}\left(\text{softmax}\left(\frac{(p_{\mathcal{V}_i} W_j^Q)(\mathcal{U}_{\mathcal{V}_i} W_j^K)^T}{\sqrt{d_{k}}}\right) \mathcal{U}_{\mathcal{V}_i} W_j^V \right)_{j=1}^m W^O
    $}
\end{equation}

where $m$ denotes the number of attention heads, $d_{p}$, $d_{\mathcal{U}}$, $d_{k}$, $d_{\mathcal{F}}$ represent the dimensions of hub node feature, topological feature, each attention head and fusion embedding, respectively. $W_j^Q \in \mathbb{R}^{d_{p} \times d_{k}}$, $W_j^K, W_j^V \in \mathbb{R}^{d_{\mathcal{U}} \times d_{k}}$ are learnable parameters for the $j$-th attention head, $W^O \in \mathbb{R}^{(m \cdot d_{k}) \times d_{\mathcal{F}}}$ is the output weight matrix.

\eetitle{Module II: Projection Network.}  We integrate query information using a contrastive learning-based two-tower model. This model projects the fusion embedding $\mathcal{F}_{\mathcal{V}_i}$ and positive/negative query samples ($\mathcal{Q}_i^+$ and $\mathcal{Q}_i^-$) into a shared latent space, minimizing the cosine similarity between hub node and positive query representations while maximizing the distance to negative samples. This objective is formalized by the following contrastive loss function:
\begin{equation}
    \label{eq:loss}
    \resizebox{0.92\linewidth}{!}{$
    \mathcal{L}(\mathcal{V}_i) = - \sum_{q \in \mathcal{Q}_i^+} \log \frac{\exp(\hat{\mathcal{F}_{\mathcal{V}_i}} \cdot \hat{p_q} / \tau)}{\sum_{q \in \mathcal{Q}_i^+} \exp(\hat{\mathcal{F}_{\mathcal{V}_i}} \cdot \hat{p_q} / \tau) + \sum_{q' \in \mathcal{Q}_i^-} \exp(\hat{\mathcal{F}_{\mathcal{V}_i}} \cdot \hat{p_{q'}} / \tau)}
    $}
\end{equation}

where $\hat{\mathcal{F}_{\mathcal{V}_i}}$ and $\hat{p_q}$ represent the normalized vectors for the hub node and query, respectively, and $\tau$ is a temperature parameter.  This contrastive loss, adapted from~\cite{chen2020simple}, encourages the fused features of hub nodes to be more similar to positive samples and less similar to negative samples. The projection network uses a lightweight \textsf{MLP} with \textsf{ReLU} activation to minimize the loss function. (see Figure~\ref{fig:pipline}).

By stitching these two modules, the Fusion Embedding Augmentation and the Projection Network, we have effectively integrated topology information and query information, generating enhanced latent representations for the hub nodes. These enhanced representations incorporate both topological semantics and query-specific features, equipping the hub nodes with the necessary information for high-tier navigation. Next, we describe how to capture the relationships between these enhanced hub nodes, enabling them to form a high-tier graph that supports adaptive navigation.

\stitle{Connecting edges between hub nodes.} Recall that we have obtained the final refined representation for each hub node, which enables the recommendation of optimal hub nodes for various queries. However, it is impractical to visit all the hub nodes each time. To further accelerate the retrieval process, we construct a similarity graph for the extracted hub nodes based on cosine similarity. Specifically, for each hub node, we connect edges to its $s$ most similar nodes based on cosine similarity, where $s \ll |\mathcal{V}|$ and $|\mathcal{V}|$ is the size of the hub nodes set. We highlight two reasons for constructing the graph in this manner: (1) This fits naturally with the model's training objective which aligns queries and hub nodes based on cosine similarity. (2) Cosine similarity satisfies the properties of a metric~\cite{sohn2016improved}, allowing for rapid convergence to the most similar (i.e., optimal) hub nodes by Algorithm~\ref{alg:gnns}.

\stitle{Summary.} \gate is composed of hub nodes and the edges between them. Our workflow gives \gate (hub nodes and edges) adaptive awareness, a flexible solution for any graph index, improving on current methods.

\subsection{Complexity Analysis}

\stitle{Indexing Time.} Let $|\mathcal{D}|$ denote the dataset size, $|\mathcal{V}|$ the number of hub nodes, $d$ the dimension, $|\mathcal{Q}|$ the historical query set size. During the preprocessing stage, hub node extraction has a time complexity of $O(k \cdot \log_k |\mathcal{D}| \cdot |\mathcal{V}| \cdot d)$, where $k$ is constant. The time complexity of the training dataset construction is $O(|\mathcal{Q}| \cdot |\mathcal{V}| \cdot S)$, where $S$ denotes the top-1 search complexity in the underlying index. In the training stage, the time complexity is $O(N \cdot L^2 \cdot d_{L})$ and the space complexity is $O(t \times (d^2 + d_{L}^2))$, where $N$ denotes training dataset size, $L$ denotes the hidden sequence length, $d_{L}$ denotes the hidden dim of model, $t$ denotes the model layers. The main bottleneck remains the construction of the NSG. 
\section{Experimental Evaluation}
\label{sec:Exp}

We conduct extensive experiments on five real-world datasets to showcase the effectiveness, efficiency, and applicability of \gate, addressing the following questions: \textbf{RQ1}: How does \gate's perform compared to competitors across varied real-world datasets? \textbf{RQ2}: What advantages does \gate gain from the proposed optimizations? \textbf{RQ3}: How sensitive is \gate to parameter variations? 

\begin{figure*}[tb!]
\vspace{1ex}
\centering
\centerline{\includegraphics[width=1\linewidth]{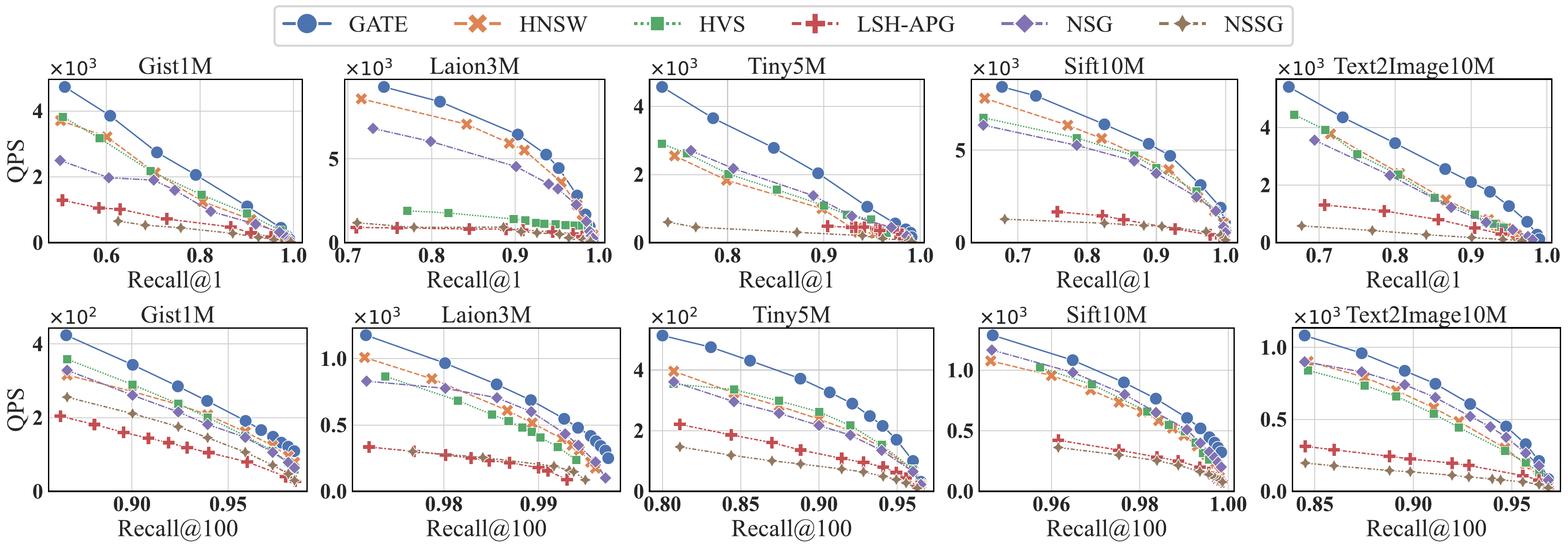}}
\vspace{-2ex}
\caption{\anns Performance on five Datasets. The top right is better.}
\vspace{-2ex}
\label{fig:rq1_topk}
\end{figure*}

\begin{table}[tb!]
  \caption{Dataset statistics. Dim indicates vector dimension.}
  \vspace{-2mm}
  \label{tab:prop}
  \resizebox{0.95\linewidth}{!}{
  \begin{tabular}{ccccc}
    \toprule
    Dataset & Base Size & Dim & Query Size& Modality\\
    \midrule
    \texttt{Gist1M}~\cite{sift_gist} & 1,000,000 & 960 & 1,000  & Image\\
    \texttt{Laion3M}~\cite{laion} & 3,000,000 & 512 & 1,000 & Multi\\
    \texttt{Tiny5M}~\cite{Tiny5M} & 5,000,000 & 384 & 1,000 & Image\\
    \texttt{Sift10M}~\cite{sift_gist} & 10,000,000 & 128 & 1,000 & Image \\
    \texttt{Text2Image10M}~\cite{text2image} & 10,000,000 & 200 & 100,000 & Multi\\

    \bottomrule
  \end{tabular}}
  \vspace{-2mm}
\end{table}

\subsection{Experimental Setup}

\stitle{Datasets.} We use five real-world datasets of varying sizes, dimensions, and modalities (shown in Table~\ref{tab:prop}), commonly used to evaluate \anns methods: \textbf{\textsf{Gist1M}}~\cite{sift_gist}, \textbf{\textsf{Tiny5M}}~\cite{Tiny5M},  \textbf{\textsf{Sift10M}}~\cite{sift_gist}, \textbf{\textsf{Laion3M}}~\cite{laion}, and \textbf{\textsf{Text2Image10M}}~\cite{text2image}.

\stitle{Competitors.} We compare \gate with five recent advanced graph-based methods: 
\textbf{(1) \textsf{HNSW}}~\cite{munoz2019hierarchical}, a state-of-the-art navigable small world graph; 
\textbf{(2) \textsf{HVS}}~\cite{lu2021hvs}, which employs a hierarchical structure of multiple \textsf{PQ} layers to reduce computational costs; 
\textbf{(3) \textsf{LSH-APG}}~\cite{zhao2023towards}, which facilitates entry point selection using a lightweight locality-sensitive hashing framework; 
\textbf{(4) \textsf{NSG}}~\cite{fu2019fast}, an approximation of the monotonic relative neighborhood graph; and 
\textbf{(5) \textsf{NSSG}}~\cite{fu2021high}, an extension of \textsf{NSG} optimized for out-of-index queries.

\stitle{Implementation.} All baselines are implemented in C++. Experiments are conducted on a CentOS machine with 128GB RAM and an Intel(R) Xeon(R) CPU E5-2650 v4 @ 2.20GHz. We use 48 threads for indexing and a single thread for query execution. Each query experiment is repeated ten times, and the average result is reported to reduce system variability. For all algorithms, we employed the grid search method to choose the relatively optimal results.

\stitle{Parameter Settings.} The parameter configurations were as follows: \textbf{(1) HNSW}: $efConstruction = 1024$, $M = 32$; \textbf{(2) HVS}: $v = 32$, $T = 6$, $efConstruction = 512$, $\delta = 0.5$; \textbf{(3) LSH-APG}: $efConstruction = 512$, $L = 2$, $p_{\tau} = 0.9$; \textbf{(4) NSG}: $L = 50$, $R = 50$, $C = 512$ (providing strong performance); \textbf{(5) NSSG}: $L = 512$, $R = 60$, $\text{Angle} = 60$; and \textbf{(6) GATE}: $|\mathcal{V}| = 512$, $h = 5$, $t_{\text{pos}} = 3$, $t_{\text{neg}} = 15$. Training employed the Adam optimizer~\cite{kingma2014adam} with a learning rate of $5 \times 10^{-5}$ for 200 epochs and a batch size of 8,192.

\stitle{Evaluation Metrics.}
We evaluate search efficiency and accuracy using Queries Per Second (QPS), a common metric for graph-based \anns~\cite{munoz2019hierarchical, dong2011efficient, jayaram2019diskann, malkov2018efficient, fu2019fast}. In single-thread scenarios, QPS also reflects latency. Additionally, we evaluate search path length (\(\ell\)) using routing hops. Let $R'$ denote the set of $k$ vectors returned by the algorithm, and $R$ represent the ground truth, the recall@k is formally defined as 
$recall@k = \frac{|R \cap R'|}{|R|} = \frac{|R \cap R'|}{k}$.

\subsection{Experimental Results}

\stitle{Exp-1: Performance Evaluation (RQ1).} We perform extensive experiments on five datasets against all competitors:

\eetitle{Overall query performance.} Figure~\ref{fig:rq1_topk} shows the queries per second (QPS) for all methods. The results highlight three key findings: \textbf{(1)} \gate outperforms all methods in query performance, achieving speedups of 2.03$\times$, 1.73$\times$, and 1.54$\times$ over \textsf{Laion3M}, \textsf{Sift10M}, and \textsf{Tiny5M}, respectively, compared to the second best competitor. \textbf{(2)} \gate exhibits strong performance at various recall@k levels, particularly at recall@1, where it effectively identifies optimal entry points to reduce search length for the top-1 nearest neighbor. \textbf{(3)} Entry point optimization techniques in \textsf{HVS} and \textsf{LSH-APG} enhance performance on \textsf{Tiny5M} and \textsf{Sift10M} but struggle on other datasets. It is because \textsf{HVS} loses information and neglects topological structure due to its multi-layer PQ-based selection, while \textsf{LSH-APG}'s effectiveness hinges on the \textsf{LSH} function's performance.

\begin{figure}[tb!]
\vspace{1ex}
\centering
\centerline{\includegraphics[width=1\linewidth]{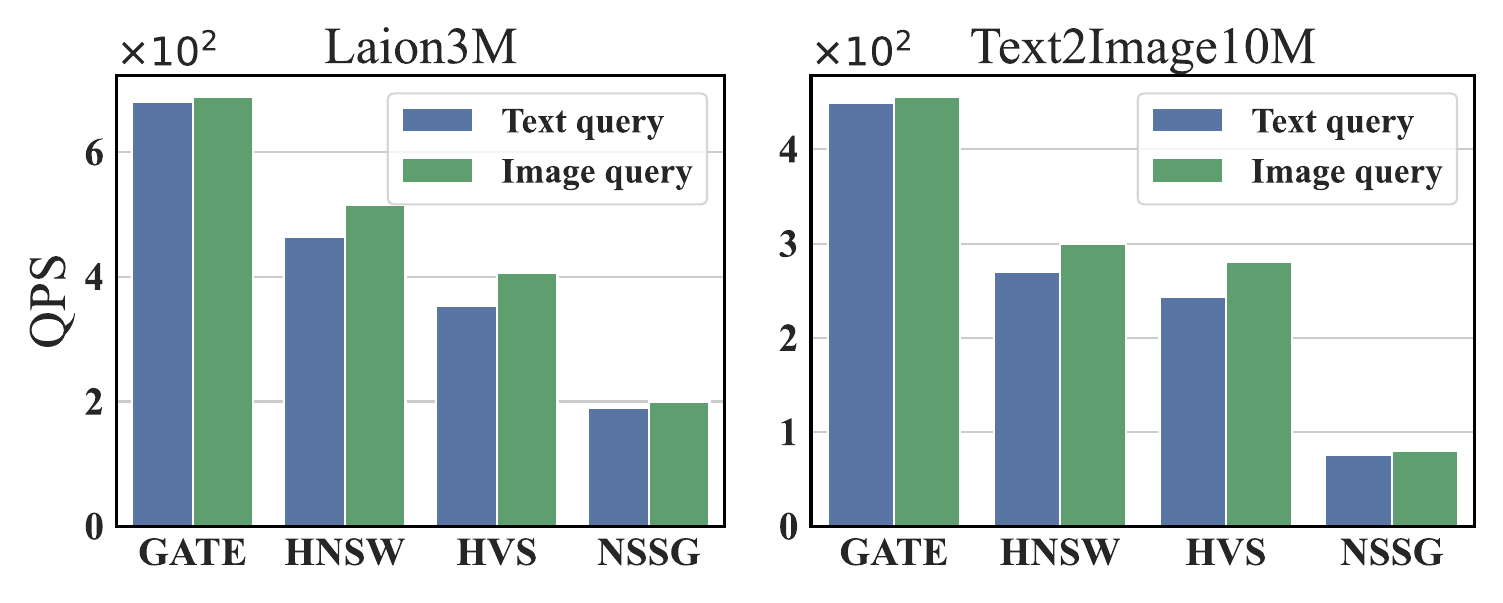}}
\vspace{-2ex}
\caption{Evaluation on different query types.}
\vspace{-3ex}
\label{fig:in_index_query}
\end{figure}

\eetitle{Robustness with out-distribution queries.} We evaluate the performance on both in- and out-distribution queries at 99\% recall level on \textsf{Laion3M} and \textsf{Text2image10M} datasets, as it is crucial for an \anns index to effectively handle different query types. Since the base data is image-based, we sample 1000 text queries and 1000 image queries respectively. The results (shown in Figure~\ref{fig:in_index_query}) indicate that \gate is both the top performer and the most robust (with only a 1.2\% gap between different types) which support our method's effectiveness. \textsf{NSSG}~\cite{fu2021high} also demonstrates good stability, but its performance is far inferior to \gate.

\eetitle{Search path analysis.} To assess search efficiency, we evaluated search path lengths on three datasets using 1000 queries. Table~\ref{tab:search_length} shows that GATE reduces the search path length by 30–40\% compared to the baseline. This behavior is consistent in the top-100 evaluation, as the graph search first approximates the top-1 result and then diffuses to identify the top-100.

\stitle{Exp-2: Ablation study (RQ2).} We evaluate the impact of key components of \gate across three datasets, examining the effects of optimizations that proposed in \S\ref{sec:methods}. Specifically, we compare five methods: (1) \gate (with all optimizations), (2) \gate without HBKM (\textbf{\gate w/o H}), (3) \gate without fusion embedding (\textbf{\gate w/o FE}), (4) \gate without contrastive learning-based loss (\textbf{\gate w/o L}), and (5) \textsf{NSG} (the underlying proximity graph of \gate). We measure the number of search hops for fair comparison at the recall@100 level of 99\% (Table~\ref{tab:ablation}). Key findings include:

\eetitle{Effectiveness of HBKM.} HBKM achieves an average improvement of 6.9\% across three datasets, thanks to its selection of representative points between clusters, which enhances inter-cluster connectivity and prevents queries from getting trapped in local optima. It achieved a 10\% improvement on \textsf{Text2image10M}, demonstrating superior performance on datasets with strong clustering properties.

\eetitle{Effectiveness of fusion embedding.} Fusion embedding adds 8\% performance improvement. It integrates the graph's topological structure into the hub node, leading to improved navigation efficiency. 

\begin{table}[tb!]
    \centering
    \caption{The search path length (\(\ell\)) comparison over various datasets at 95\% recall.}
    \vspace{-2mm}
    \label{tab:search_length}
    \resizebox{\linewidth}{!}{
    \begin{tabular}{c|cccc}
        \toprule
        Search Path Length & \begin{tabular}[c]{@{}c@{}}Gist1M\\ (Recall@1=95)\end{tabular} & \begin{tabular}[c]{@{}c@{}}Tiny5M\\ (Recall@1=95)\end{tabular} & \begin{tabular}[c]{@{}c@{}}Text2Image10M\\ (Recall@1=95)\end{tabular} &  \\
        \midrule
        NSG                & 5236                & 7334                & 10181                      &  \\
        HVS                & 6283                & 9021                & 12235                      &  \\
        \midrule
        GATE               & 3824                & 4914                & 6818                       & \\
        \bottomrule
    \end{tabular}
    }
    \vspace{-4mm}
\end{table}

\eetitle{Effectiveness of contrastive learning-based loss.} This loss is an indispensable component, as it is intrinsically linked to the training of the two-tower model. Cosine similarity contributes to stable and effective \gate training. Our ablation study shows that without this loss, performance drops by approximately 30\%.

\stitle{Exp-3: Parameter Sensitivity (RQ3).} We assess the effects of parameter tuning on \textsf{Sift10M}, focusing on two key parameters: (1) the subgraph sample hop $h$ and (2) the query sample threshold $t_{pos}$.

\begin{figure}[tb!]
    \vspace{1ex}
    \centering
    \centerline{\includegraphics[width=1\linewidth]{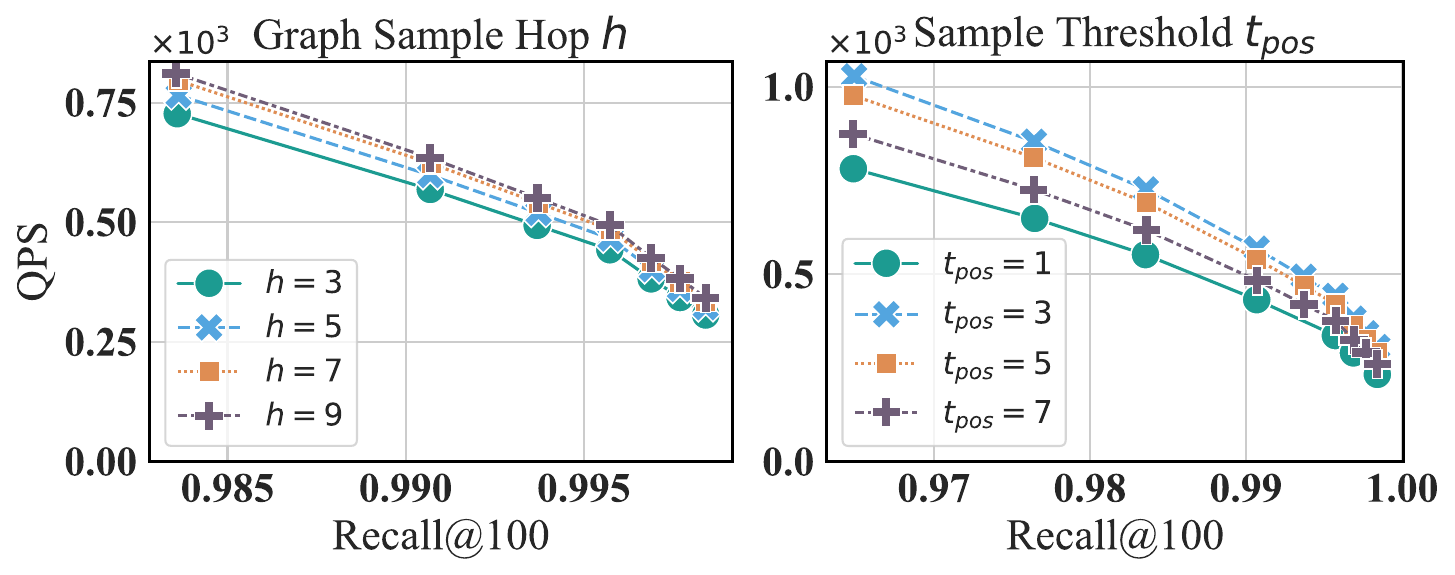}}
    \vspace{-2ex}
    \caption{Performance on \textsf{Sift10M} with tuned parameters.}
    \label{fig:rq3}
    \vspace{-1.8ex}
\end{figure}

\eetitle{Impact of subgraph sample hops $h$.} We vary the max subgraph sample hop $h \in \{3, 5, 7, 9 \}$. Figure ~\ref{fig:rq3} demonstrates that (1) the query performance of \gate positively correlates with increasing $h$, as higher $h$ capture richer graph features (it also increase the time cost); (2) the performance gains gradually diminish because excessively large $h$ result in homogenized graph features across nodes.

\eetitle{Impact of query sample threshold $t_{pos}$.} We vary $t_{pos} \in \{1, 3, 5, 7\}$, which influences the likelihood of a query being classified as a positive sample. Figure~\ref{fig:rq3} illustrates that the optimal $t_{pos}$ is convex ($t_{pos}=3$ is the best), simplifying parameter tuning across different datasets (a similar trend is observed for $t_{neg}$, but is omitted due to space limitation).

\section{Related Work}
\label{sec:related} 

\stitle{Approximate Nearest Neighbor Search}. Approximate Nearest Neighbor Search (\anns) in high-dimensional spaces is vital for recommendation systems~\cite{covington2016deep,grbovic2018real,yi2019sampling}, information retrieval~\cite{petitjean2014dynamic}, and large language models~\cite{achiam2023gpt,lewis2020retrieval}. \anns methods are broadly divided into: (1) Tree-based~\cite{silpa2008optimised,beckmann1990r}; (2) \textsf{LSH}-based~\cite{gionis1999similarity,datar2004locality,huang2015query,liu2014sk}; (3) Quantization-based~\cite{ge2013optimized,jegou2010product,norouzi2013cartesian,zhang2019grip}; and (4) Graph-based~\cite{munoz2019hierarchical,dong2011efficient,jayaram2019diskann,malkov2018efficient}. This paper focuses on graph-based methods for their superior \anns performance.

\stitle{Graph-based Methods.} Graph-based methods are inspired by early paradigms with solid theoretical foundations, such as Delaunay Graphs (\textsf{DG}), Relative Neighborhood Graphs (\textsf{RNG}), k-Nearest Neighbor Graphs (\textsf{KNNG}), and \textsf{MSNET}~\cite{aurenhammer1991voronoi, dearholt1988monotonic, toussaint1980relative}. However, constructing these theoretical-sounding proximity graph in high-dimensional spaces incurs high time complexity. Consequently, several approximate indexing strategies are proposed to reduce the indexing complexity~\cite{munoz2019hierarchical, dong2011efficient, jayaram2019diskann, malkov2018efficient, fu2019fast, fu2016efanna, fu2021high, harwood2016fanng, li2019approximate,chen2024roargraph, chen2025maximum, chen2025stitching}.

\stitle{Two-Tower Model.} The two-tower model, a deep learning architecture with dual neural networks, has gained popularity in tasks like text retrieval~\cite{gillick2018end}, response suggestion~\cite{hu2008collaborative}, and recommendation systems~\cite{yi2019sampling,huang2019fibinet}. Each tower, constructed upon a representation network, aligns embeddings via sampled positives and negatives. In recommendation systems, the left and right towers encode user and item features, respectively, enabling real-time user embedding inference and offline item embedding precomputation for efficient similarity-based recommendations. Our work highlights the two-tower model's effectiveness in constructing an \anns index.

\begin{table}[tb!]
    \centering
    \caption{Ablation study results. We use search hops to measure the benefits of three proposed methods.}
    \vspace{-2mm}
    \label{tab:ablation}
    \resizebox{\linewidth}{!}{
    \begin{tabular}{c|ccc}
        \toprule
        Methods & Gist1M & Sift10M & Text2Image10M \\ 
        \midrule
        \gate & 7566(+18.2\%) & 7415(+25.5\%) & 10686(+30.3\%) \\ 
        \gate w/o H & 7990(+13.7\%) & 8139(+18.3\%) & 12080(+21.2\%) \\ 
        \gate w/o FE & 8091(+12.6\%) & 7976(+19.6\%) & 11425(+25.5\%) \\ 
        \gate w/o L & 11424(-23.3\%) & 11270(-13.1\%) & 14752(+3.9\%) \\ 
        \midrule
        NSG & 9260 & 9966 & 15343 \\
        \bottomrule
    \end{tabular}
    }
    \vspace{-4mm}
\end{table}

\section{Conclusion}
\label{sec:conclude}
This paper presents \gate, a high-tier proximity graph that improves the query performance of graph indexes by linking \anns with recommendation systems. Our methods include (i) a hierarchical balanced k-means approach for hub node selection; (ii) a subgraph and query sampling technique to provide hub nodes with topology and query information; and (iii) a contrastive learning-based two-tower model to empower \gate with adaptive awareness. Extensive experiments demonstrate the effectiveness of our techniques in overcoming the limitations of existing graph-based \anns. In the future, we shall consider more diverse large-scale datasets and develop more efficient training techniques.

\begin{acks}
Jiancheng Ruan, Tingyang Chen, Xiangyu Ke are supported in part by CCF-Aliyun2024004, Ningbo Yongjiang Talent Introduction Programme (2022A-237-G). Yunjun Gao is supported in part by the NSFC under Grants No. (62025206 and U23A20296), Zhejiang Province’s "Lingyan" R\&D Project under Grant No. 2024C01259. Renchi Yang is supported by the National Natural Science Foundation of China (No. 62302414), the Hong Kong RGC ECS grant (No. 22202623), and the Huawei Gift Fund.
\end{acks}

\bibliographystyle{ACM-Reference-Format}
\balance
\bibliography{ref}






\end{document}